\documentclass[prl,preprint,amsmath,amssymb]{revtex4}
\usepackage{graphicx}
\usepackage{booktabs}
\usepackage{threeparttable}
\usepackage[colorlinks,
            linkcolor=blue,
            anchorcolor=blue,
            citecolor=blue]{hyperref}
\newtheorem{theorem}{Theorem}

\def\ra{\rangle}
\def\la{\langle}

\allowdisplaybreaks[4]

\begin{document}

\title{A note on uncertainty relations of arbitrary $N$ quantum channels}

\author{Qing-Hua Zhang$^{1,}$\footnotemark[1]}
\author{Jing-Feng Wu$^{1,}$\footnotemark[1]}
\author{Shao-Ming Fei$^{1,2,}$\footnotemark[1]}

\affiliation{$^1$School of Mathematical Sciences, Capital Normal University,
Beijing 100048, China\\
$^2$Max-Planck-Institute for Mathematics in the Sciences, 04103 Leipzig, Germany}

\renewcommand{\thefootnote}{\fnsymbol{footnote}}

\footnotetext[1]{Corresponding authors. \\
\href{mailto:2190501022@cnu.edu.cn}{2190501022@cnu.edu.cn(Q. H. Zhang)}.\\
\href{mailto:2180502107@cnu.edu.cn}{2180502107@cnu.edu.cn(J. F. Wu)}.\\
\href{mailto:feishm@cnu.edu.cn}{feishm@cnu.edu.cn(S. M. Fei)}.}
\bigskip

\begin{abstract}
Uncertainty principle plays a vital role in quantum physics. The Wigner-Yanase skew information characterizes the uncertainty of an observable with respect to the measured state. We generalize the uncertainty relations for two quantum channels to arbitrary $N$ quantum channels based on Wigner-Yanase skew information. We illustrate that these uncertainty inequalities are tighter than the existing ones by detailed examples. Especially, we also discuss the uncertainty relations for $N$ unitary channels, which could be regarded as variance-based sum uncertainty relations with respect to any pure state.
\end{abstract}

\maketitle

\section{\expandafter{\romannumeral1}. INTRODUCTION}
As a fundamental characteristic of quantum theory, uncertainty principle has been widespread concerned since Heisenberg proposed the notions of uncertainties for measuring non-commuting observables \cite{WH1927}. The well-known Robertson uncertainty relation says that for arbitrary two observables $A$ and $B$ \cite{HR1929},
$\Delta A\Delta B\geq \frac{1}{2}|\la \psi [A,B]\psi\ra|$, where the commutator $[A,B]=AB-BA$ and $\Delta \Omega=\sqrt{\la \Omega^2\ra-\la\Omega\ra^2}$ is the standard deviation of an observable $\Omega$ with respect to the measured state $|\psi\ra$. With the development of quantum information theory, many kinds of characterizations and quantifications of uncertainty relations have been established, such as the ones based on entropy \cite{HMJU1988, LRZP2014, VNPA2016, DD1983, SWSY2009, AR2013, PCMB2017, FASS2020}, Wigner-Yanase skew information \cite{SL2003,BCSFGL2016,LZTG2021}, under successive measurements \cite{MS2003,JDSP2013,KBTF2014,JZYZ2015,BCSF2015A}, and with majorization techniques \cite{IBLR2011,ZPLR2013,SFVG2013}.

In modern formalism of quantum theory, the most general description of quantum measurement is given in terms of quantum channels \cite{MNIC2000,PBMGPL1997}. Quantum channels play a pivotal role in quantum information processing. Many aspects related to the quantum channels have been extensively investigated, such as the coherence of quantum channels \cite{VCIG2019,TTDZ2018}, the operational resource theory of quantum channels \cite{LLKB2018,YL2020}, the capacity of quantum channels \cite{SL1997,FCVG2014}, and the abilities of quantum channels in producing or destroying quantum resources \cite{KB2001,SLSF2010,FGFP2013,LZZM2018}. Recently, the uncertainty relations for quantum channels have been also widely studied in terms of the Wigner-Yanase skew information and variance \cite{SL2003,JRJB2009,YSNL2021,LZTG2021}. The Wigner-Yanase skew information $I_{\rho}(A)$ with respect to a quantum state $\rho$ and an arbitrary operator $A$ is defined by \cite{SLYZ2019,YF2018, SLYS2018,EWMY1963},
\begin{equation}
I_{\rho}(A)=\frac{1}{2}tr([\sqrt{\rho},A]^\dagger[\sqrt{\rho},A])=\frac{1}{2}\| [\sqrt{\rho},A]\|^2,
\end{equation}
where $\|\bullet \|$ denotes the Frobenius norm. Let $\Phi$ be a quantum channel with Kraus representation, $\Phi(\rho)=\sum_{i=1}^{n}K_i\rho K_i^\dagger$. The Wigner-Yanase skew information of $\rho$ with respect to the channel is given by
\begin{equation}
I_{\rho}(\Phi)=\sum_{i=1}^{n}I_{\rho}(K_i),
\end{equation}
where $I_{\rho}(K_i)=\frac{1}{2}Tr([\sqrt{\rho},K_i]^\dagger[\sqrt{\rho},K_i])$\cite{SLYS2018}. The quantity $I_{\rho}(\Phi)$ is well-defined because it is independent of the choice of the Kraus of $\Phi$. It is demonstrated that $I_{\rho}(\Phi)$ can be regarded as a bona fide measure, for coherence as well as quantum uncertainty of $\rho$ with respect to quantum channel $\Phi$.  

For a pure state, the skew information for channel has a similar physical meaning to the variance \cite{APUS2015}. Pass through a given quantum channel $\Phi$ with Kraus operators $K_i$, a pure state $|\psi\ra$ can be transformed into:
$$\rho=\Phi(\rho)=\sum_{i=1}^{n}K_i |\psi \ra\la \psi | K_i^\dagger.$$
The Fidelity between $|\psi\ra$ and $\rho$ is defined as \cite{MNIC2000}:
$$F=\la \psi|\rho|\psi\ra=\sum_{i=1}^{n}|\la \psi|K_i |\psi\ra|^2.$$
The skew information of $|\psi\ra$ for quantum channel $\Phi$
$$I_{|\psi\ra}(\Phi)=\sum_i I_{|\psi\ra}(K_i)=1-\sum_{i=1}^{n}|\la \psi|K_i |\psi\ra|^2=1-F,$$
that is to say, $$I_{|\psi\ra}(\Phi)+F=1.$$
The above equality shows a strict complementarity between fidelity and uncertainty of quantum channel \cite{LZTG2021}. The complementary relation reveals that any restriction on the uncertainty in the channel will impose a restriction on the fidelity between the input and output states.

Fu $et\ al.$ established the uncertainty relation for two quantum channels $\Phi_1$ and $\Phi_2$ in terms of Wigner-Yanase skew information \cite{SFYS2019},
\begin{equation}
I_{\rho}(\Phi_1)+I_{\rho}(\Phi_2) \geq \max_{\pi \in S_n}\frac{1}{2}\sum_{i=1}^n I_{\rho}(K_i^1\pm K_{\pi(i)}^2),
\end{equation}
where $\Phi_s=\sum_{i=1}^nK_i^s \rho (K_i^s)^\dagger$, $s=1,2$, $\pi \in S_n$ is an arbitrary $n$-element permutation.

Very recently, generalizing the results in \cite{SFYS2019} to the case of $N$ quantum channels, Zhang $et\ al.$ provided two elegant uncertainty relations \cite{LZTG2021},
\begin{equation}
\begin{aligned}\label{zhangeq1}
\sum_{s=1}^{N}I_{\rho}(\Phi_s) \geq \max_{\pi_s,\pi_t \in S_n}
&\frac{1}{N-2}\sum_{i=1}^n\Bigg \{\sum_{1\leq s<t\leq N} I_{\rho}(K_{\pi_s(i)}^s+K_{\pi_t(i)}^t) \\
&-\frac{1}{(N-1)^2}\Bigg [\sum_{1\leq s<t\leq N}\sqrt{ I_{\rho}(K_{\pi_s(i)}^s+K_{\pi_t(i)}^t)} \Bigg ]^2\Bigg \},
\end{aligned}
\end{equation}
and
\begin{equation}\label{zhangeq2}
\begin{aligned}
\sum_{s=1}^{N}I_{\rho}(\Phi_s) \geq \max_{\pi_s,\pi_t \in S_n}
&\frac{1}{N}\sum_{i=1}^n \Bigg \{ I_{\rho}(\sum_s K_{\pi_s(i)}^s) \\
&+ \frac{2}{N(N-1)}\Bigg [\sum_{1\leq s<t\leq N}\sqrt{ I_{\rho}(K_{\pi_s(i)}^s-K_{\pi_t(i)}^t)} \Bigg ]^2\Bigg \},
\end{aligned}
\end{equation}
where $\pi_s,\pi_t \in S_n$ are arbitrary $n$-element permutations. For convenience, we denote the right hands of (\ref{zhangeq1}) and  (\ref{zhangeq2}) as $\overline{\rm LB}_1, \overline{\rm LB}_2$, respectively.

In this paper, we formulate several new uncertainty relations based on Wigner-Yanase skew information for $N$ quantum channels. The lower bounds of our uncertainty inequalities are tighter than the existing ones \cite{LZTG2021}. Detailed examples are presented to illustrate the superiority. Especially, we also discuss the uncertainty relations for unitary channels. 

\section{\expandafter{\romannumeral2}. Skew information-based sum uncertainty relations for quantum channels}
Let $\Phi$ be a quantum channel with Kraus representation, $\Phi(\rho)=\sum_{i=1}^{n}K_i\rho K_i^\dagger$. 
The skew information of the channel can be written as
\begin{equation} \label{s2eq2}
I_{\rho}(\Phi)=\frac{1}{2}tr(a^\dagger a) =\frac{1}{2}\|a\|^2,
\end{equation}
where $a^\dagger=([\sqrt{\rho},K_1]^\dagger, [\sqrt{\rho},K_2]^\dagger, \dots, [\sqrt{\rho},K_n]^\dagger)$. $I_{\rho}(\Phi)$ characterizes some intrinsic features of both the quantum state and the quantum channel. For arbitrary $N$ quantum channels, we have the following conclusion.

\begin{theorem}\label{th1}
Let $\Phi_1,\Phi_2,\dots,\Phi_N$ be $N$ quantum channels with Kraus representations $\Phi_s(\rho)=\sum_{i=1}^{n}K_i^s\rho (K_i^s)^\dagger$, $s=1,2,...,N$. We have
\begin{equation}\label{th1eq0}
\sum_{s=1}^{N}I_{\rho}(\Phi_s) \geq {\rm Max }\{{\rm LB1},{\rm  LB2}, {\rm LB3}\},
\end{equation}
where
\begin{equation}
\begin{aligned}\label{th1eq1}
{\rm LB1}= \max_{\pi_s,\pi_t \in S_n}
&\frac{1}{N-2}\Bigg \{\sum_{1\leq s<t\leq N}\sum_{i=1}^n I_{\rho}(K_{\pi_s(i)}^s+K_{\pi_t(i)}^t) \\
&-\frac{1}{(N-1)^2}\Bigg [\sum_{1\leq s<t\leq N}\sqrt{\sum_{i=1}^n I_{\rho}(K_{\pi_s(i)}^s+K_{\pi_t(i)}^t)} \Bigg ]^2\Bigg \},
\end{aligned}
\end{equation}
\begin{equation}\label{th2eq1}
\begin{aligned}
{\rm LB2}= \max_{\pi_s,\pi_t \in S_n}
&\frac{1}{N}\Bigg \{\sum_{i=1}^n I_{\rho}(\sum_{s=1}^N K_{\pi_s(i)}^s) \\
&+ \frac{2}{N(N-1)}\Bigg [\sum_{1\leq s<t\leq N}\sqrt{\sum_{i=1}^n I_{\rho}(K_{\pi_s(i)}^s-K_{\pi_t(i)}^t)} \Bigg ]^2\Bigg \},
\end{aligned}
\end{equation}
\begin{equation}\label{th3eq1}
\begin{aligned}
{\rm LB3}= \max_{\pi_s,\pi_t \in S_n}
&\frac{1}{2N-2}\Bigg \{\sum_{1\leq s<t\leq N}\sum_{i=1}^n I_{\rho}(K_{\pi_s(i)}^s\mp K_{\pi_t(i)}^t) \\
&+ \frac{2}{N(N-1)}\Bigg [\sum_{1\leq s<t\leq N}\sqrt{\sum_{i=1}^n I_{\rho}(K_{\pi_s(i)}^s\pm K_{\pi_t(i)}^t)} \Bigg ]^2\Bigg \},
\end{aligned}
\end{equation}
$\pi_s,\pi_t \in S_n$ are arbitrary $n$-element permutations.
\end{theorem}

{\sf [Proof]} To prove the inequality (\ref{th1eq0}), we employ the following equality,
$$
\|\sum_{s=1}^N a_s\|^2+(N-2)\sum_{s=1}^N \|a_s\|^2=\sum_{1\leq s<t \leq N}\|a_s+a_t\|^2.
$$
Note that
$$\|\sum_{s=1}^N a_s\|=\|\frac{1}{N-1}\sum_{1\leq s<t \leq N}(a_s+a_t)\| \leq \frac{1}{N-1}\sum_{1\leq s<t \leq N}\|a_s+a_t\|,
$$
we get
\begin{equation*}
\sum_{s=1}^N\|a_s\|^2 \geq \frac{1}{N-2} [\sum_{1\leq s<t\leq N}\|a_s+a_t\|^2-\frac{1}{(N-1)^2}(\sum_{1\leq s<t\leq N}\|a_s+a_t\|)^2].
\end{equation*}
From (\ref{s2eq2}), we have $\|a_s\|^2=2I_{\rho}(\Phi_s)$ and $\|a_s+a_t\|^2=2\sum_{i=1}^N I_{\rho}(K_i^s+K_i^t)$, which proves $\sum_{s=1}^{N}I_{\rho}(\Phi_s) \geq {\rm LB1}$.

By using the identity,
$$
N\sum_{s=1}^N \|a_s\|^2=\|\sum_{s=1}^N a_s\|^2+\sum_{1\leq s<t \leq N}\|a_s-a_t\|^2,
$$
and the Cauchy-Schwarz inequality, we obtain
$$
\sum_{1\leq s<t \leq N}\|a_s-a_t\|^2 \geq \frac{2}{N(N-1)}(\sum_{1\leq s<t \leq N}\|a_s-a_t\|)^2,
$$
and
$$
\sum_{s=1}^N \|a_s\|^2 \geq \frac{1}{N}[\|\sum_{s=1}^N a_s\|^2+\frac{2}{N(N-1)}(\sum_{1\leq s<t \leq N}\|a_s-a_t\|)^2].
$$
Taking account into that $\|\sum_{s=1}^N a_s\|^2=2\sum_{i=1}^n I_{\rho}(\sum_s K_{i}^s)$ and $\|a_s - a_t\|^2=2\sum_{i=1}^N I_{\rho}(K_i^s - K_i^t)$, we prove the inequality $\sum_{s=1}^{N}I_{\rho}(\Phi_s) \geq {\rm LB2}$.

At last, by using the parallelogram law,
$$
(2N-2)\sum_{s=1}^N \|a_s\|^2=\sum_{1\leq s<t\leq N}\|a_s-a_t\|^2 + \sum_{1\leq s<t\leq N}\|a_s+a_t\|^2
$$
and the Cauchy-Schwarz inequality, we get
$$
\sum_{1\leq s<t \leq N}\|a_s - a_t\|^2 \geq \frac{2}{N(N-1)}(\sum_{1\leq s<t \leq N}\|a_s - a_t\|)^2
$$
and
$$
\sum_{1\leq s<t \leq N}\|a_s + a_t\|^2 \geq \frac{2}{N(N-1)}(\sum_{1\leq s<t \leq N}\|a_s + a_t\|)^2.
$$
Therefore we have
$$
\sum_{s=1}^{N} \| a_s\|^2 \geq \frac{1}{2N-2}[\frac{2}{N(N-1)}(\sum_{1\leq s<t \leq N} \| a_s \mp a_t \|)^2 + \sum_{1\leq s<t \leq N} \| a_s\pm a_t \|^2],
$$
which proves the inequality $\sum_{s=1}^{N}I_{\rho}(\Phi_s) \geq {\rm LB3}$. $\Box$

As examples, let us consider the mixed state given by Bloch vector $\vec{r}=(\frac{\sqrt{3}}{2}cos\theta,\frac{\sqrt{3}}{2}sin\theta,0)$ \cite{LZTG2021},
\begin{equation}                 \label{ex1}
\rho=\frac{I_2+\vec{r}\cdot\vec{\sigma}}{2},
\end{equation}
where $\vec{\sigma}=(\sigma_x,\sigma_y,\sigma_z)$ is given by the standard Pauli matrices, $I_2$ is the $2\times 2$ identity matrix. We respectively consider three quantum channels: the phase damping channel $\phi$,
$$
\phi(\rho)=\sum_{i=1}^2A_i\rho (A_i)^\dagger,\quad A_1=|0\ra\la0|+\sqrt{1-q}|1\ra\la1|,\quad A_2=\sqrt{q}|1\ra\la1|,
$$
the amplitude damping channel $\epsilon$,
$$\epsilon(\rho)=\sum_{i=1}^2B_i\rho (B_i)^\dagger,\quad B_1=|0\ra\la0|+\sqrt{1-q}|1\ra\la1|,\quad B_2=\sqrt{q}|0\ra\la1|
$$
and the bit flip channel $\Lambda$,
$$
\Lambda(\rho)=\sum_{i=1}^2C_i\rho (C_i)^\dagger,\quad C_1=\sqrt{q}|0\ra\la0|+\sqrt{q}|1\ra\la1|,\quad C_2=\sqrt{1-q}(|0\ra\la1|+|1\ra\la0|)
$$
with $0\leq q<1$.

\begin{figure}[tbp]
 \centering
 \includegraphics[width=11cm]{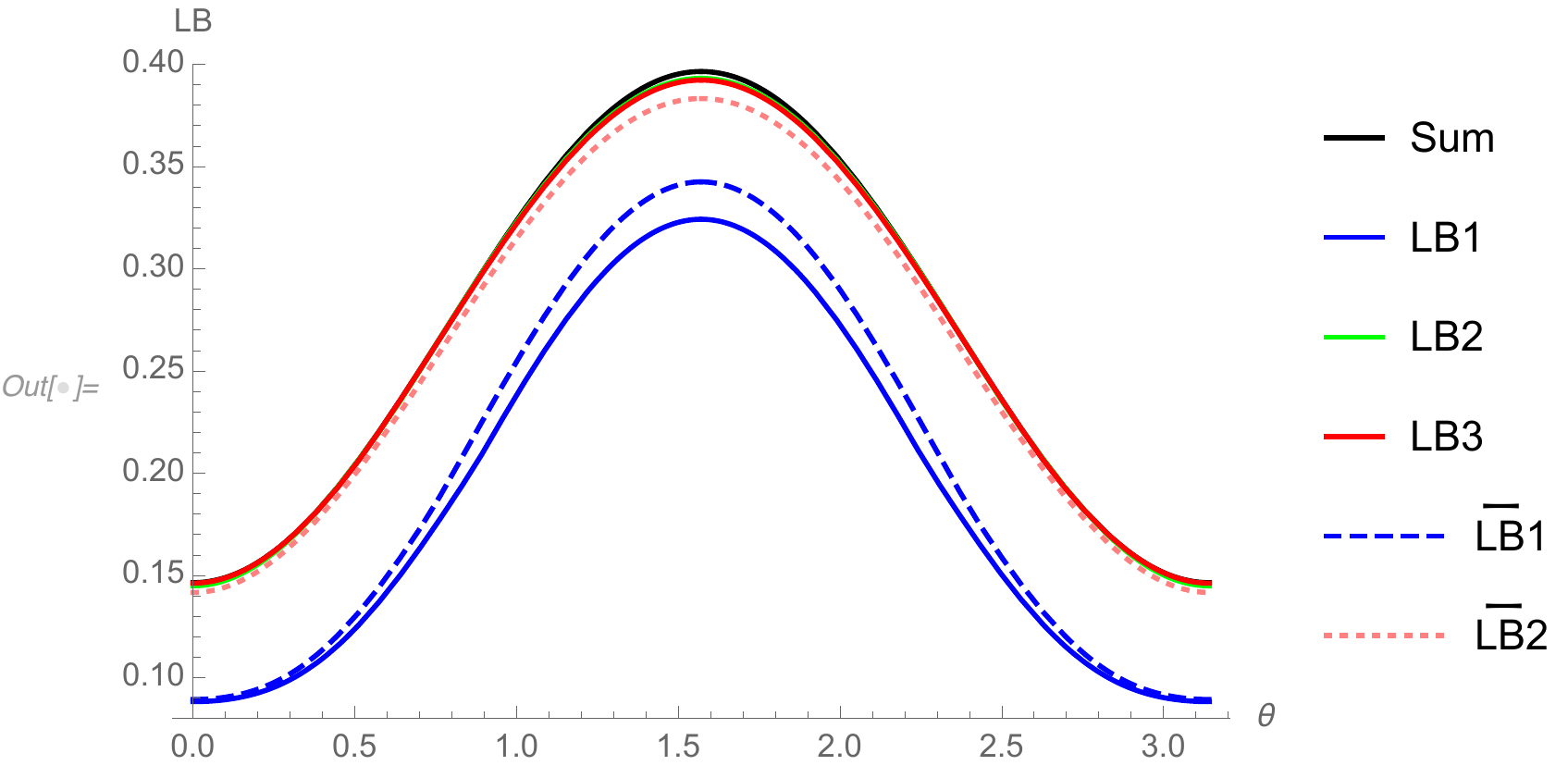}
 \caption{The comparison among the lower bounds $\overline{\rm LB}1$, $\overline{\rm LB}2$, ${\rm LB1}$, ${\rm LB2}$ and ${\rm LB3}$ for the state $\rho$ with Bloch vector $\vec{r}=(\frac{\sqrt{3}}{2}cos\theta,\frac{\sqrt{3}}{2}sin\theta,0)$, and three quantum channels, the phase damping channe, the amplitude damping channel and the bit flip channel. Let ${\rm Sum}=I_{\rho}(\phi)+I_{\rho}(\epsilon)+I_{\rho}(\Lambda)$.}
  \label{figex1}
 \end{figure}
 
For the case $q=0.1$ and $\theta=\pi/2$, we have $I_{\rho}(\phi)+I_{\rho}(\epsilon)+I_{\rho}(\Lambda)=0.475658$. The lower bound LB1 is 0.449135, the lower bounds $\overline{\rm LB}1$ and $\overline{\rm LB}2$ are  0.475658 and 425827, respectively. Obviously, ${\rm LB}1$ is tighter than $\overline{\rm LB}2$ in \cite{LZTG2021}. Here the lower bound LB1 is also greater than 0.42873 from LB2 and 0.440462 from LB3. 

We also consider the case $q=0.5$, the comparison among the lower bounds $\overline{\rm LB}1$, $\overline{\rm LB}2$, LB1, LB2 and LB3 is shown in Figure. \ref{figex1}. Especially, we take some special $\theta$, see Table. \ref{tab1}. These results show that our Theorem 1 improve the existing ones given in \cite{LZTG2021}.

\begin{table}[htbp]
\caption{Comparison among the uncertainty lower bounds}\label{tab1}
\begin{tabular*}{\hsize}{@{}@{\extracolsep{\fill}}lllllll@{}}
		\hline
$q=0.5$&$\overline{\rm LB}1$&$\overline{\rm LB}2$&LB1
&LB2&LB3&$I_{\rho}(\phi)+I_{\rho}(\epsilon)+I_{\rho}(\Lambda)$ \\
		\hline		
		$\theta=\pi/6$& 0.133979& 0.204181&0.127677&0.208898 &0.20891 & 0.208947 \\
		$\theta=\pi/4$& 0.194803& 0.264726 &0.182753&0.271447 & 0.271447 & 0.271447 \\
		$\theta=\pi/2$&0.342466&0.383224  &0.324177&0.393068 &0.393913 & 0.396447 \\
		\hline
\end{tabular*}
\end{table}

In addition, by using the generalized Hlawka's inequality \cite{AHYO1998,BCSF2015B,BCSFGL2016},
$$
\sum_{s=1}^N \|a_s\| \geq \frac{1}{N-2}(\sum_{1\leq s<t \leq N}\| a_s+a_t\| - \|\sum_{i=1}^N a_s\|),
$$
we can similarly prove the following theorem.

\begin{theorem}\label{th4}
Let $\Phi_1,\Phi_2,\dots,\Phi_N$ be $N$ quantum channels with Kraus representations $\Phi_s(\rho)=\sum_{i=1}^{n}K_i^s\rho (K_i^s)^\dagger$, we have
\begin{equation}\label{th4eq1}
\sum_{s=1}^{N}\sqrt{I_{\rho}(\Phi_s)} \geq \max_{\pi_s,\pi_t \in S_n}
\frac{1}{N-2}\Bigg \{\sum_{1\leq s<t\leq N} \sqrt{\sum_{i=1}^n I_{\rho}(K_{\pi_s(i)}^s+ K_{\pi_t(i)}^t)} - \sqrt{\sum_{i=1}^n I_{\rho}(\sum_{s=1}^N K_{\pi_s(i)}^s)} \Bigg \},
\end{equation}
where $\pi_s,\pi_t \in S_n$ are arbitrary $n$-element permutations.
\end{theorem}

Unitary channels are also used a lot in quantum computation and quantum information theory\cite{MNIC2000}. Consider an arbitrary channel $U(\rho)=U\rho U^\dagger$, the Wigner-Yanase skew information of $\rho$ with respect to the channel is given by
$$I_{\rho}(U)=\frac{1}{2}tr([\sqrt{\rho},U]^\dagger[\sqrt{\rho},U]).$$

Next we consider the skew information-based uncertainty relations for $N$ unitary channels $U_1, U_2, \dots,U_N$.  Directly from Theorom 1, the following uncertainty relations hold: 
\begin{equation}\label{th5eq1}
\sum_{s=1}^{N}I_{\rho}(U_s) \geq 
\frac{1}{N-2}\Bigg \{\sum_{1\leq s<t\leq N} I_{\rho}(U_{s}+U_{t}) -\frac{1}{(N-1)^2}\Bigg [\sum_{1\leq s<t\leq N}\sqrt{I_{\rho}(U_{s}+U_{t}}) \Bigg ]^2\Bigg \},
\end{equation}
\begin{equation}\label{th5eq2}
\sum_{s=1}^{N}I_{\rho}(U_s) \geq 
\frac{1}{N}\Bigg \{I_{\rho}(\sum_{s=1}^N U_{s}) 
+ \frac{2}{N(N-1)}\Bigg [\sum_{1\leq s<t\leq N}\sqrt{I_{\rho}(U_{s}-U_{t})} \Bigg ]^2\Bigg \},
\end{equation}
\begin{equation}\label{th5eq3}
\sum_{s=1}^{N}I_{\rho}(U_s) \geq 
\frac{1}{2N-2}\Bigg \{\sum_{1\leq s<t\leq N} I_{\rho}(U_{s}\mp U_{t}) 
+ \frac{2}{N(N-1)}\Bigg [\sum_{1\leq s<t\leq N}\sqrt{I_{\rho}(U_{s}\pm U_{t}}) \Bigg ]^2\Bigg \}.
\end{equation} 
For convenience, we denote the right hands of (\ref{th5eq1}), (\ref{th5eq2}) and (\ref{th5eq3}) as Lb1, Lb2 and Lb3, respectively.

Theorem 2 implies the following inequality holds for $N$ Unitary channels:
\begin{equation}\label{th6eq1}
\sum_{s=1}^{N}\sqrt{I_{\rho}(U_s)} \geq 
\frac{1}{N-2}\Bigg \{\sum_{1\leq s<t\leq N} \sqrt{ I_{\rho}(U_{s}+ U_{t})} - \sqrt{ I_{\rho}(\sum_{s=1}^N U_{s})} \Bigg \}.
\end{equation}
Noticed that quantum variance is defined as: $(\Delta_{|\psi\ra}U)^2=\frac{1}{2}\la UU^\dagger+U^\dagger U\ra-\la U\ra\la U^\dagger\ra$ with any quantum pure state $|\psi\ra$, then the following equality holds:
\begin{equation}
(\Delta_{|\psi\ra}U)^2=I_{|\psi\ra}(U).
\end{equation}
The above inequalities (\ref{th5eq1}) to (\ref{th6eq1}) can be regarded as variance-based sum uncertainty relations for $N$ unitary operators.

\begin{figure}[htbp]
 \centering
 \includegraphics[width=11cm]{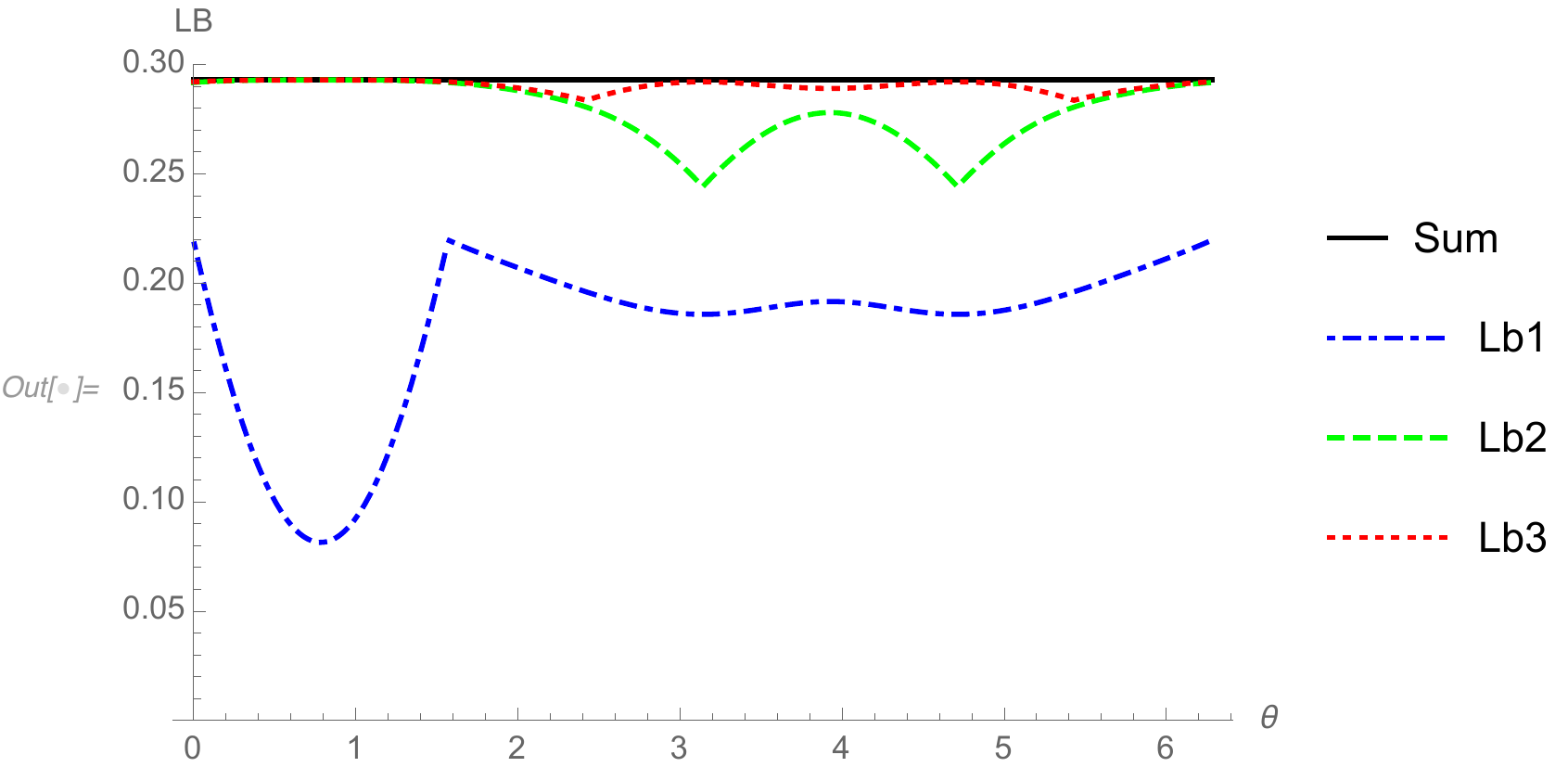}
 \caption{The black (solid) line is ${\rm Sum}=I_{\rho}(U_1)+I_{\rho}(U_2)+I_{\rho}(U_3)=1-\sqrt{2}/2$. The blue (dot-dashed) line Lb1, green (dashed) line Lb2 and red (dotted) line Lb3 represent the right hands of (\ref{th5eq1}), (\ref{th5eq2}) and  (\ref{th5eq3}), respectively.}
  \label{figex2}
 \end{figure}

We take an example to illustrate these uncertainty relations. Let us consider the pure state $\rho=\frac{1}{2}(I+\vec{r}\cdot\vec{\sigma})$ with $\vec{r}=(\frac{1}{\sqrt{2}}cos\theta,\frac{1}{\sqrt{2}}sin\theta,\frac{1}{\sqrt{2}})$, where $\sigma_x$,$\sigma_y$,$\sigma_z$ are Pauli matrices.

Consider three unitary operators,
\begin{equation*}
\begin{gathered}
U_1=e^{\frac{i\pi \sigma_x}{8}}=
\begin{pmatrix} cos\frac{\pi}{8} & i sin\frac{\pi}{8} \\ i sin\frac{\pi}{8} &cos\frac{\pi}{8} \end{pmatrix},\\
U_2=e^{\frac{i\pi \sigma_y}{8}}=
\begin{pmatrix} cos\frac{\pi}{8}  & sin\frac{\pi}{8} \\ -sin\frac{\pi}{8}  & cos\frac{\pi}{8}  \end{pmatrix},\\
U_3=e^{\frac{i\pi \sigma_z}{8}}=
\begin{pmatrix} e^{i \frac{\pi}{8}} & 0 \\ 0&e^{-i\frac{\pi}{8}} \end{pmatrix},
\end{gathered}
\end{equation*}
which correspond to Bloch sphere rotations of $-\pi/4$ about the x axis, the y axis and z axis, respectively. Then the lower bounds of inequalities (\ref{th5eq1}), (\ref{th5eq2}) and (\ref{th5eq3}) associated with $\rho$ can be computed. Figure \ref{figex2} shows that  lower bound of (\ref{th5eq3}) is strictly greater than  (\ref{th5eq1}) and   (\ref{th5eq2}) in this case.

\section{\expandafter{\romannumeral3}. CONCLUSION}
Based on Wigner-Yanase skew information for quantum channels, we have derived several uncertainty relations for arbitrary $N$ quantum channels. By detailed examples we have shown that our uncertainty relations improve the existing ones. We also get several uncertainty relations for $N$ unitary channels. It can be regarded as variance-based sum uncertainty relations for $N$ unitary operators as we take the pure state. These results and the simple approaches used in this article may highlight further investigations on related uncertainty relations.

\bigskip
\noindent{\bf Acknowledgments}\, This work is supported by NSFC (Grant No. 12075159), Key Project of Beijing Municipal Commission of Education (KZ201810028042), Beijing Natural Science Foundation (Z190005), Academy for Multidisciplinary Studies, Capital Normal University, the Academician Innovation Platform of Hainan Province, and Shenzhen Institute for Quantum Science and Engineering, Southern University of Science and Technology (No. SIQSE202001).

\end{document}